\documentclass[twocolumn,superscriptaddress,showpacs,amsmath,amssymb,aps,pra,floatfix]{revtex4-1}
\usepackage{graphicx}
\usepackage{bm}
\usepackage{amssymb}
\usepackage{amsmath}
\usepackage{amsthm}
\usepackage{amsfonts}
\usepackage{mathrsfs}
\usepackage[colorlinks=true,citecolor=blue,urlcolor=black]{hyperref}

\usepackage{times}

\begin{document}

\title{Experimental asymmetric Plug-and-Play Measurement-device-independent quantum key distribution}

\author{Guang-Zhao Tang}\affiliation{College of Science, National University of Defense Technology, Changsha 410073, People's Republic of China}

\author{Shi-Hai Sun}\email{shsun@nudt.edu.cn}\affiliation{College of Science, National University of Defense Technology, Changsha 410073, People's Republic of China}

\author{Feihu Xu} \email{fhxu@mit.edu.cn}
\affiliation{Research Laboratory of Electronics, Massachusetts Institute of Technology, 77 Massachusetts Avenue, Cambridge, Massachusetts 02139, USA}

\author{Huan Chen}\affiliation{College of Science, National University of Defense Technology, Changsha 410073, People's Republic of China}

\author{Chun-Yan Li}\affiliation{College of Science, National University of Defense Technology, Changsha 410073, People's Republic of China}

\author{Lin-Mei Liang} \email{nmliang@nudt.edu.cn}
\affiliation{College of Science, National University of Defense Technology, Changsha 410073, People's Republic of China}
\affiliation{State Key Laboratory of High Performance Computing, National University of Defense Technology, Changsha 410073, People's Republic of China}

\date{\today}

\begin{abstract}
Measurement-device-independent quantum key distribution (MDI-QKD) is immune to all security loopholes on detection. Previous experiments on MDI-QKD required spatially separated signal lasers and complicated stabilization systems. In this paper, we perform a proof-of-principle experimental demonstration of plug-and-play MDI-QKD over an asymmetric channel setting with a single signal laser, in which the whole system is automatically stabilized in spectrum, polarization, arrival time and phase reference. Both the signal laser and the single-photon detectors are in the possession of a common server. A passive timing calibration technique is applied to ensure the precise and stable overlap of signal pulses. The results pave the way for the realization of a quantum network, in which the users only need the encoding devices.
\end{abstract}

\maketitle

\section{Introduction}
Quantum key distribution (QKD) \cite{BB84,E91} allows the two legitimate parties, Alice and Bob, to share an information-theoretical secure key guaranteed by the laws of quantum physics. Despite tremendous experimental efforts being made in the field~\cite{Yuan08,Liu10,Wang12}, an important problem in current QKD implementations is the gap between its theory
and practice~\cite{Gisin06,Xu10,Lydersen10,Sun11,Jain11}. To close this gap, three main approaches have been developed. The first one is the security patch~\cite{Yuan10,Silva12}, but it is difficult to include all potential and unnoticed security loopholes. The second one is the device-independent QKD (DI-QKD)~\cite{Acin07,Gisin2010,Curty2011}. This approach is still challenging with current technology, since it requires a loophole-free Bell test. The third approach is the measurement-device-independent QKD (MDI-QKD)~\cite{Lo12}, which removes all detection-related security loopholes. Such loophole is arguably the most important issue identified in conventional QKD implementations~\cite{Xu10,Lydersen10,Sun11,Jain11}. Therefore, MDI-QKD is of great importance to promote the security of practical QKD systems. In addition, with current technology, MDI-QKD is suitable for both long distance communication and metropolitan networks~\cite{Xu15,Tang2016}.

Achievements of MDI-QKD have been made in both theory~\cite{Ma12,Xu14,Wang14,Curty14,XuNP15} and experiment \cite{Rubenok13,Liu13,Tang14,TangL14,Ferreira13,Wang15,Comandar16,Tang2016}. The experimental demonstration of MDI-QKD requires the indistinguishability of photons from Alice and Bob, mainly in three dimensions: spectrum, polarization, and timing. To solve this challenge, the active stabilization systems were normally utilized in previous experiments. For example, the feedback temperature-control units for the distributed feedback lasers~\cite{Liu13,TangL14} or frequency-locked lasers \cite{Rubenok13,Tang14} were employed to match the spectral mode. The feedback temporal-control system was utilized to calibrate the arrival time of signals~\cite{Rubenok13,TangL14}. The phase (or polarization) stabilization system was always essential in the time-bin phase-encoding (or polarization-encoding) scheme. Recently, many proposals and demonstrations have been made to mitigate the complexity in the implementations of MDI-QKD~\cite{Gonzalez15,Liang2014,Xu15,Choi16}. In particular, a promising scheme is the plug-and-play MDI-QKD~\cite{Xu15,Choi16}, which greatly reduces the experimental complexity of mode matching and reference-frame alignment. But, since the signal laser source and single photon detectors (SPDs) are in the charge of an untrusted server, plug-and-play MDI-QKD is vulnerable to source attacks~\cite{Gisin06,Xu10,Sun11}. Fortunately, with the security analysis reported in~\cite{Xu15}, plug-and-play MDI-QKD can be implemented even with a single untrusted source. However, so far, an experimental demonstration of plug-and-play MDI-QKD is still missing, except for a proof-of-concept test~\cite{Choi16} with polarization encoding over free space channel within a few meters. This proof-of-concept test was operated in a wavelength outside the window of telecom-wavelength, and even so, it still needs some stabilization measures.

In this paper, we report a demonstration of plug-and-play MDI-QKD using time-bin phase-encoding over an asymmetric channel setting, in which the two channels from Alice and Bob to Charlie are 14 km and 22 km standard optical fibers. The encoding optical pulses of Alice and Bob come from a single homemade laser held by Charlie, which ensures that no mismatch exists in both pulse waveform and optical spectrum. Thanks to the plug-and-play architecture~\cite{Stucki02}, the polarization state is automatically calibrated and stabilized. The encoding optical pulses of Alice and Bob share the same reference frame. In the asymmetric channel setting, an experimental challenge is how to precisely match the timing of pulses, returned respectively from Alice and Bob over two mismatched channels. We developed a passive timing calibration method by using two synchronization lasers (operating at 1310 nm) and multiplexing them with the signal laser (operating at 1550 nm) via wavelength division multiplexing (WDM). As we know, this is the first time that the passive timing calibration method is applied in the experiment of MDI-QKD.

\begin{figure*}[htbp]
  \centering
  \includegraphics[width=0.6\linewidth]{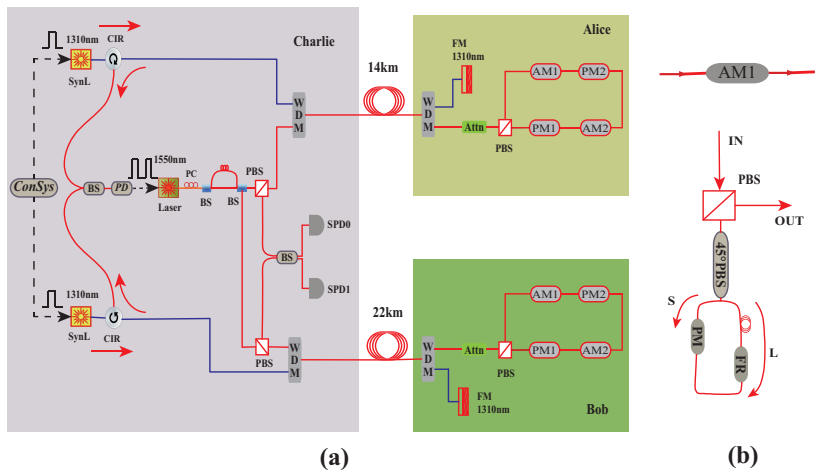}
  \caption{(a) Experimental setup of the plug-and-play MDI-QKD. The synchronization optical pulses (1310nm) are sent out by Charlie. They travel to Alice (Bob), and are reflected back by a Faraday mirror (1310nm). A homemade photoelectric detector (PD) is utilized to detect them and output the system clock. Then, the signal laser (1550nm) generates the optical pulses of Bob (Alice). The time bins are generated by an AMZI in Charlie. The clients (Alice and Bob) receive the time-bin pulses and encode the bits. Then, the signals are reflected back to Charlie for the partial BSM. ConSys, control system; SynL, synchronization laser; CIR, circulator; BS, beam splitter; PC, polarization controller; WDM, wavelength division multiplexer; Attn, attenuator; FM, Faraday mirror; SPD, single-photon detector. (b) Schematic of the amplitude modulator (AM1): PBS, polarizing beam splitter; $45^{0}$ PBS, polarizing beam splitter with $45^{0}$ from the optical axis; PM, phase modulator; FR, Faraday rotator.} ~\label{Fig:setup}
\end{figure*}

\section{Experiment}\label{sec:Experiment}
The experimental setup of plug-and-play MDI-QKD is illustrated in Figure~\ref{Fig:setup}(a). We implement the time-bin phase-encoding scheme~\cite{Rubenok13,Liu13,TangL14,Wang15,Tang2016}. The signal laser source (1550nm) and detectors are held by a common server (Charlie). The signal laser is internally modulated into a pulse train with a width of 2 ns (FWHM) at 1 MHz repetition rate. An asymmetric Mach-Zehnder interferometer (AMZI) is utilized to separate the pulses into two time bins with 20 ns time delay. Alice and Bob only have the modulation devices including phase modulators (PMs) and amplitude modulators (AMs) for encoding. The key bit in $X$ basis is encoded into the relative phase, 0 or $\pi$, by PM1, while the key bit in $Z$ basis is encoded into the time bin, 0 or 1, by AM1. Figure~\ref{Fig:setup}(b) illustrates the structure of AM1~\cite{Martinez09}. AM1 contains a normal polarizing beam splitter (PBS), a PBS with $45^{\circ}$ rotation from the optical axis, a phase modulator, and a Faraday rotator, in which different optical intensities at the output port of PBS are realized by modulating the relative phase between the long path (L) and the short path (S)~\cite{CommentAM1}. PM2 is used for the active phase randomization, and AM2 is used to implement the decoy states.

The signal pulses travel through a single mode fiber spool of 14km (22km) from Charlie to Alice (Bob). After being modulated by Alice and Bob, the pulses return to Charlie interfere at the beam splitter (BS). A coincident detection at alternative time bins indicates a successful Bell state measurement (BSM). At the measurement site, the BSM is implemented with a polarization-maintaining BS and two commercial InGaAs SPDs (ID201) with an efficiency of $10\%$ and a gate width of 2.5 ns. The dead time is $10$ $\mu s$ with a dark count rate of $6\times10^{-6}$ per gate. A Bell state (Singlet) is successfully post-selected when a coincidence of two SPDs happens at alternative time bins. After Charlie announces the result of the partial BSM, Alice and Bob sift the raw key.

The crucial aspect in the experiment is the indistinguishability of signal pulses from Alice and Bob, mainly in three dimensions: spectrum, polarization, and timing. Errors would occur if any mismatch exists in these dimensions. In our system, Alice and Bob share the same signal laser, which guarantees no mismatch in spectrum and in pulse waveform. The active phase randomization is implemented to eliminate the partial-phase-randomization attack. In our proof-of-principle demonstration, a sawtooth wave with a repetition rate of 55 KHz (15 KHz) is applied to the PM2 of Alice (Bob)~\cite{CommentPM2}, to globally randomize the phase of each optical pulse in the range of [0,2$\pi$]. Alice's and Bob's time bins come from the same AMZI, so they share the same phase reference frame. For the polarization mode, the plug-and-play architecture can automatically compensate for the birefringence effects~\cite{Stucki02}.

% We can denote the polarization state of optical pulses sent by Charlie as horizontal polarization. After being modulated by Alice (Bob), they travel back to Charlie and result in vertical polarization. Therefore, the polarization state is perfectly aligned and stabilized.

In the asymmetric channel setting, the pulses of Alice and Bob travel different lengths of fibers. A challenge is to match the temporal mode. We use two additional synchronization lasers (SynL, 1310 nm) to calibrate the arrival time. The whole system is synchronized in the following manner: the SynL pulses are sent from Charlie to Alice (Bob), reflected back by a Faraday mirror (1310 nm), and detected by a photoelectric detector (PD). The output of PD is used to drive the signal laser (1550nm) to generate the signal pulses of Bob (Alice). The temporal mode difference between Alice and Bob can be expressed as:
\begin{equation}
\begin{aligned}
\Delta t=&(t_{C\rightleftarrows B}^{1310}+t_{C\rightleftarrows A}^{1550})-(t_{C\rightleftarrows A}^{1310}+t_{C\rightleftarrows B}^{1550}) \\
=&\Delta t_{0}+(1/v_{1550}-1/v_{1310})\Delta L
\end{aligned}
\end{equation}
where $\Delta t_{0}=(1/v_{1550}-1/v_{1310})(L_{C\rightleftarrows B}^{0}-L_{C\rightleftarrows A}^{0})$, and $\Delta L=\Delta L_{C\rightleftarrows B}-\Delta L_{C\rightleftarrows A}$. $L_{C\rightarrow B}$ represents the fiber length between Charlie and Bob. $\Delta L=\alpha_{T}L^{0}\Delta T$,
where $\alpha_{T}=5.4\times10^{-7}/^{\circ}C$ is the thermal expansion coefficient of fiber, and $\Delta T$ represents the change of temperature. The second term in Eq. (1) is negligible since it only induces 0.14 ps with $10^{\circ}C$ temperature change~\cite{CommentdeltaT}. Therefore, the arrival time difference of signals between Alice and Bob is a constant which can be compensated by adjusting the time delay between two SynLs with a delay chip. The temporal mode mismatch depends on the resolution of the delay chip (10ps) which is much smaller than the width of the signal pulse (2ns). This ensures a high-visibility interference.

\section{Results}\label{sec:Results}
In our demonstration, the optical pulses are modulated into three different intensities according to the decoy state method~\cite{Xu13}, namely signal state intensity ($\mu=0.4$), decoy state intensity ($\nu=0.1$) and vacuum state intensity ($\omega=0.01$). The optical intensities of a certain basis are put into 9 pairs. The experimental gains and quantum bit error rates (QBER) for different intensity combinations are listed in Table~\ref{Tab:results1} and Table~\ref{Tab:results2}. The QBERs of Z-basis are due to the extinction ratio of AM1 and the background counts (Rayleigh backscattering and detectors' dark counts). In the ideal case, the QBERs of Z-basis should be 0. While, in X-basis, the vacuum and multiphoton components of weak coherent states cause accidental coincidences which introduce an error rate of $50\%$. Thus, the error rate of X-basis has an expected value of $25\%$.

The secure key is extracted from the data when both Alice and Bob encode their bits using signal states ($\mu$) in the Z basis. The rest data is applied to estimate the parameters used in the secure key rate calculation. In the asymptotic case, the secure key rate is given by~\cite{Lo12}
\begin{equation}
R\geq q\{Q_{\mu\mu,11}^{Z,L}[1-H(e_{11}^{X,U})]-Q_{\mu\mu}^{Z}fH(E_{\mu\mu}^{Z})\},
\end{equation}
where $q$, $Q_{\mu\mu}^{Z}$, and $E_{\mu\mu}^{Z}$ are the possibility, overall gain, and QBER when Alice and Bob send the signal states in the Z basis. $Q_{\mu\mu,11}^{Z,L}=\mu^{2}e^{-2\mu}Y_{11}^{Z,L}$, where $Y_{11}^{Z,L}$ is a lower bound of the yield of single photon states in the $Z$ basis; $e_{11}^{X,U}$ is an upper bound of the QBER of the single photon states in the $X$ basis; $Y_{11}^{Z,L}$ and $e_{11}^{X,U}$ can be estimated from the decoy state method; $f$ is the error correction efficiency; $H$ is the binary Shannon entropy function. A total number of $N=6.14\times10^{10}$ pulses are sent out in the experiment. We take the value $q=\frac{1}{18}$ and $f=1.16$ in our calculation. By using the analytical bounds derived in~\cite{Xu13}, we obtain that $Y_{11}^{Z,L}= 2.2\times10^{-3}$ and $e_{\mu\mu,11}^{X,U}=5.07\%$ (see Appendix~\ref{App}). Finally, a secure key rate of $R=4.7\times10^{-6}$ bits per pulse is demonstrated.

\begin{table}[htp]
\centering\caption{Experimental values of gains $Q_{I_{A}I_{B}}^{Z(X)}$($\times10^{-4}$). $I_{A}$ and $I_{B}$ are the optical intensities of Alice and Bob.}
\begin{tabular}{lp{0.5in} lp{0.5in} lp{0.5in} lp{0.5in} lp{0.5in} lp{0.5in}} \hline
 & & Z-basis & & & X-basis \\ \hline
 & $I_{A}$ & & & $I_{B}$ \\ \hline
 & $\mu$ & $\nu$ & $\omega$ & $\mu$ & $\nu$ & $\omega$  \\ \hline
$\mu $ & 1.819 & 0.547 & 0.125 & 9.018 & 4.347 & 3.408  \\
$\nu $  & 0.624 & 0.217 & 0.0378 & 4.316 & 0.925 & 0.323   \\
$\omega $ & 0.130 & 0.0386 & 0.0050 & 5.207 & 0.323 & 0.0115   \\ \hline
\end{tabular}  \label{Tab:results1}
\end{table}

\begin{table}[h!]
\centering\caption{Experimental values of QBERs. Error bars represent one standard deviation.}
\begin{tabular}{lp{0.6in} lp{0.6in} lp{0.6in} lp{0.6in} lp{0.6in} lp{0.6in}} \hline
 & & Z-basis & & & X-basis \\ \hline
 & $I_{A}$ & & & $I_{B}$ \\ \hline
& $\mu$ & $\nu$ & $\omega$ & $\mu$ & $\nu$ & $\omega$  \\ \hline
$\mu $ & 0.0188 & 0.0378 & 0.136 & 0.269 & 0.341 & 0.483  \\
       & $\pm 0.001$ & $\pm0.004$ & $\pm0.009$ & $\pm0.007$ & $\pm0.007$ & $\pm0.009$ \\
$\nu $  & 0.0356 & 0.0450 & 0.133 & 0.351 & 0.278 & 0.428   \\
       & $\pm 0.003$ & $\pm0.003$ & $\pm0.013$ & $\pm0.008$ & $\pm0.012$ & $\pm0.012$ \\
$\omega $ & 0.151 & 0.133 & 0.194 & 0.484 & 0.432 & 0.368   \\
       & $\pm 0.005$ & $\pm0.01$ & $\pm0.04$ & $\pm0.008$ & $\pm0.015$ & $\pm0.052$ \\ \hline
\end{tabular} \label{Tab:results2}
\end{table}

\section{Conclusion and discussion}\label{sec:Conclusion}
We discuss the limitations of our proof-of-principle experiment and possible solutions. First, the plug-and-play MDI-QKD is vulnerable to various source attacks~\cite{Gisin06,Xu10,Sun11}, since the signal source is totally controlled by an untrusted server. With the assumptions including the single mode assumption, the phase randomization assumption, and the trust of the monitoring devices, a complete security analysis for plug-and-play MDI-QKD was derived in~\cite{Xu15}. It shows that, with careful source monitoring, we can rigorously derive a lower bound of the secure key generation rate even with an unknown and untrusted source. According to the security analysis in~\cite{Xu15}, the energy and arrival time of each signal pulse should be monitored precisely to acquire the certain information about the photon-number distribution and the timing mode. In our demonstration, however, we cut down the energy of signal pulses to reduce the Rayleigh backscattering (RBS), which lead that the intensity detector can't monitor such weak energy. Hence, the monitor unit was not implemented. This can be improved by using the scheme of pulse trains, as demonstrated in conventional plug-and-play QKD~\cite{Stucki02,Sun10}. Second, in our implementation, the parameters were not optimized and the secure key rate was only calculated in the asymptotic case. A full parameter optimization and the finite-key effect can be considered by using the theory in~\cite{Curty14}. Lastly, the secure key generation rate can be significantly improved by increasing the repetition rate and the detector efficiency~\cite{TangL14}.

In conclusion, we have performed a proof-of-principle demonstration of self-stabilized asymmetric plug-and-play MDI-QKD over 36 km fiber. The homemade laser sources and expensive detectors are provided by a common server. The polarization and phase can be automatically calibrated and stabilized. The passive time calibration technique ensures a precise and stable interference of photons from two remote parties. The techniques demonstrated in our experiment greatly improve the practicability of MDI-QKD and pave the way for a MDI quantum network with an untrusted network server.

\begin{acknowledgments}
We thank V. Makarov for the helpful discussions. This work is supported by the National Natural Science Foundation of China, Grant No. 11304391, 11674397. L.M.L is supported by the program of NCET. F. X. thanks NSERC PDF for the support.
\end{acknowledgments}

\appendix
\section{Secure key rate estimation} \label{App}
%The secure key rate is calculated with an analytical method with two decoy states according to \cite{Xu13}. In our calculation, we assume the three optical intensity states are prepared with the same probability. Thus, we choose $q=\frac{1}{18}$. The main wok is to estimate $Y_{11}^{Z,L}$ and $e_{11}^{X,U}$. The gains ($Q_{I_{A}I_{B}}^{Z(X)}$) are listed in Table \uppercase\expandafter{\romannumeral2}.
%

The secure key rate is calculated with an analytical method with two decoy states according to~\cite{Xu13}. $Y_{11}^{Z,L}$ is given by
\begin{equation}
\begin{aligned}
&Y_{11}^{Z,L}=                                                \\ &\frac{(\mu_{a}^{2}-\omega_{a}^{2})(\mu_{b}-\omega_{b})Q_{Z}^{M1}-(\nu_{a}^{2}-\omega_{a}^{2})(\nu_{b}-\omega_{b})Q_{Z}^{M2}}{(\mu_{a}-\omega_{a})(\mu_{b}-\omega_{b})(\nu_{a}-\omega_{a})(\nu_{b}-\omega_{b})(\mu_{a}-\nu_{a})}
\end{aligned}
\end{equation}
where $Q_{Z}^{M1}=Q_{Z}^{\nu_{a}\nu_{b}}e^{(\nu_{a}+\nu_{b})}+Q_{Z}^{\omega_{a}\omega_{b}}e^{(\omega_{a}+\omega_{b})}-Q_{Z}^{\nu_{a}\omega_{b}}e^{(\nu_{a}+\omega_{b})}-Q_{Z}^{\omega_{a}\nu_{b}}e^{(\omega_{a}+\nu_{b})}$, $Q_{Z}^{M2}=Q_{Z}^{\mu_{a}\mu_{b}}e^{(\mu_{a}+\mu_{b})}+Q_{Z}^{\omega_{a}\omega_{b}}e^{(\omega_{a}+\omega_{b})}-Q_{Z}^{\mu_{a}\omega_{b}}e^{(\mu_{a}+\omega_{b})}-Q_{Z}^{\omega_{a}\mu_{b}}e^{(\omega_{a}+\mu_{b})}$.

$e_{11}^{X,U}$ is
\begin{equation}
\begin{aligned}
e_{11}^{X,U}=&\frac{1}{(\nu_{a}-\omega_{a})(\nu_{b}-\omega_{b})Y_{11}^{X,L}}  \\
&[Q_{X}^{\nu_{a}\nu_{b}}E_{X}^{\nu_{a}\nu_{b}}e^{(\nu_{a}+\nu_{b})}+Q_{X}^{\omega_{a}\omega_{b}}E_{X}^{\omega_{a}\omega_{b}}e^{(\omega_{a}+\omega_{b})}- \\
&Q_{X}^{\nu_{a}\omega_{b}}E_{X}^{\nu_{a}\omega_{b}}e^{(\nu_{a}+\omega_{b})}-Q_{X}^{\omega_{a}\nu_{b}}E_{X}^{\omega_{a}\nu_{b}}e^{(\omega_{a}+\nu_{b})}]
\end{aligned}
\end{equation}
where $Y_{11}^{X,L}$ can be achieved with a similar method to $Y_{11}^{Z,L}$.

By using the above equations, we estimate the parameters listed in Table~\ref{Tab:parameters}. In the secure key calculation, we assume an error correction code with $f=1.16$, and choose $q=\frac{1}{18}$, which is due to the fact that the three optical intensity states are prepared with the same probability.

\begin{table}[htp]
\centering\caption{Parameters estimated in the process of secure key rate estimation. $Q_{\lambda}^{M1(2)}$($10^{-4}$), $Y_{11}^{\lambda,L}$($10^{-3}$), with $\lambda\in\{X, Z\}$.}
\begin{tabular}{lp{0.5in} lp{0.8in} lp{0.8in} lp{0.8in}} \hline
$       $ & $Q_{\lambda}^{M1}$ & $Q_{\lambda}^{M2}$ & $Y_{11}^{\lambda,L}$   \\ \hline
$Z$ & 0.1846  & 3.668 & 2.219      \\
$X$  & 0.4353 & 10.016 & 4.40     \\  \hline
\end{tabular} \label{Tab:parameters}
\end{table}

%In the experiment, the real detection efficiencies of two SPDs are different, even though we set them both to $10\%$. The optical intensities we modulate is not strictly equal to the values we want. For example, the signal state intensity is always equal to 0.41 or 0.42, and the decoy state intensity is always equal to 0.11 or 0.12. The parameters we estimated are listed in Table \uppercase\expandafter{\romannumeral3}.

%%%%%%%%%%%%%%%%%%%%%%%%%%%%%%%%%%%%%%%%%%%%%%%%%%%%%%%%%%%%%%%%%%%%%%%%%%%%%%%%%%%%%%%%%%%%
% Create the reference section
%


\begin{thebibliography}{99}
\bibitem{CommentAM1} In the experiment, we only have two traditional amplitude modulators without the bias control circuits. Thus, we use the settings of Fig. 1(b) to serve as amplitude modulators. This kind of amplitude modulator features an outstanding stability which has been demonstrated in~\cite{Martinez09}. To be honest, this method would bring in an additional optical pulse at the input port of PBS, which would cause the leakage of bit information. But it doesn't matter since we can eliminate this problem simply by replacing it with a traditional AM.
\bibitem{CommentPM2} In our demonstration, for simplicity, we take advantage of the simplest active phase randomization method. The phase is not uniformly distributed in the range of [0,2$\pi$]. But, it doesn't affect the feasibility of our demonstration, since the effective phase-randomization technique has been demonstrated in several experiments.
\bibitem{CommentdeltaT} Drift in the fiber length is affected by the temperature change and the mechanical vibration. In the experiment, the mechanical vibration plays less important role than the temperature change. Therefore, we only take the effect of temperature change into consideration.
\end{thebibliography}
\end{document}